\documentclass[11pt]{article}
\usepackage{hyperref}
\pdfoutput=1
\begin{document}
\title{Hydrodynamics and evaporation of a sessile drop of capillary size}
\author{L.Yu. Barash \\
\\ \vspace{6pt} Landau Institute for Theoretical Physics \\ 142432 Chernogolovka, Russia}
\maketitle
\begin{abstract}
Fluid dynamics video of an evaporating sessile drop of capillary size is presented.
The corresponding simulation represents the description taking into account jointly
time dependent hydrodynamics, vapor diffusion and thermal conduction in an evaporating
sessile drop. The fluid convection in the drop is driven by Marangoni forces associated
with the temperature dependence of the surface tension.
For the first time the evolution of the vortex structure
in the drop during an evaporation process is obtained.
\end{abstract}

\section{Introduction}

Evaporation of a drop in an ambient gas was considered since Maxwell
time mainly as diffusion of vapor from a near-surface
layer. On the other hand, fluid flows inside a sessile drop
driven by thermocapillary forces during the evaporation process
have been studied in detail only recently, even in
the quasistationary limit.
We have carried out for the first time
the numerical simultaions that provide 
time dependent fluid convection,
vapor diffusion and temperature distribution in an axially symmetrical 
evaporating sessile drop of capillary size~\cite{c1,c3,c2}. 
Thus, our results allow, in particular, to observe the fluid dynamics
including the time evolution of the vortex structure of Marangoni convection
as it develops in an evaporating sessile drop.

The particular conditions and parameters of the simulation
correspond to experimental conditions of evaporating toluene drops 
on silicon nitride substrate described in~\cite{c1}.
Due to high thermal conduction of silicon nitride,
the boundary condition for the temperature distribution at the substrate can
be reduced to the constant room temperature.
The vapor diffusion outside the drop results in
inhomogeneous mass flow from the drop surface, which in turn results
in the inhomogeneous temperature distribution in the drop and on the drop
surface. Marangoni forces associated with the temperature-dependent surface
tension induce the fluid flows inside the viscous drop.

In the linked video, we have used the simulation described in~\cite{c1,c3}
to study the evaporation process of a sessile drop of capillary size.
Several dynamical stages of the Marangoni convection
of an evaporating sessile drop are obtained.
The stages are characterized by different number
of vortices in the drop and the spatial location of vortices.
As seen in the video, during the early stage the array of vortices arises near
a surface of the drop and induces a non-monotonic spatial
distribution of the temperature over the drop surface.
The number of near-surface vortices in the drop
is controlled by the Marangoni cell size, which is calculated
similar to that given by Pearson for flat fluid layers.
The number of vortices quickly decreases with time,
resulting in three bulk vortices in the intermediate stage.
The vortex structure finally evolves into the single convection vortex
in the drop, existing during about $1/2$ of the evaporation time.

\section{The video}

The video showing the evaporating sessile drop can be seen at the following URLs:
\begin{itemize}
\item \href{anc/droplet-LO.mpg}{Video 1} -- Low resolution
\item \href{anc/droplet-HI.mpg}{Video 2} -- High resolution
\end{itemize}

This video has been submitted to the {\it Gallery of Fluid Motion 2010} which is an annual
showcase of fluid dynamics videos.

\end{document}